# Spin transport in a lateral spin valve with a suspended Cu channel

Kenjiro Matsuki,[1,†] Ryo Ohshima,[1,†] Livio Leiva,[1] Yuichiro Ando,[1]

Teruya Shinjo,[1] Toshiyuki Tsuchiya,[2] and Masashi Shiraishi[1*]

1. Department of Electronic Science and Engineering, Kyoto Univ., 615-8510 Kyoto, Japan.
2. Department of Micro Engineering, Kyoto Univ., 615-8540 Kyoto, Japan,

[†]Authors contributed equally

[*]Corresponding author

We study spin transport through a suspended Cu channel by an electrical non-local 4-terminal measurement for future spin mechanics applications. A magnetoresistance due to spin transport through the suspended Cu channel is observed, and its magnitude is comparable to that of a conventional fixed Cu lateral spin valve. The spin diffusion length in the suspended Cu channel is estimated to be 340 nm at room temperature from the spin signal dependence on the distance between the ferromagnetic injector and detector electrodes. This value is found to be slightly shorter than in a fixed Cu. The decrease in the spin diffusion length in the suspended Cu channel is attributed to an increase in spin scattering originating from naturally oxidized Cu at the bottom of the Cu channel.

**Introduction**

Spin mechanics, a coupling between spin angular momentum and mechanical motion, has emerged as one of the modern spintronics fields[1]. A typical example is the Einstein-de Hass effect, where reorientation of the magnetization induces mechanical rotation due to angular-momentum conservation.[2,3] It has not been easy to study this effect because the spin torque created by this effect is minute relative to the total angular momentum in a material, though many efforts have been made to do so. However, a recent study updated the situation by using spin-wave spin currents to inject a sufficient amount of angular momentum into a ferrimagnetic insulator.[4] The authors fabricated an $Y_3Fe_5O_{12}$ (YIG)-based cantilever and generated a thermal gradient along the long axis direction of the YIG cantilever, which allowed generating a spin-wave spin current in the YIG by the spin-Seebeck effect. When a resonance vibration of the YIG cantilever was detected using a laser Doppler vibrometer, an additional resonance vibration that appeared at different frequencies was observed under the spin current injection into the YIG, which is ascribed to the generation of mechanical motion by an additional magnetization under the spin current injection, i.e., manifestation of the Einstein-de Haas effect.

Towards further progress of the Einstein-de Haas effect from a modern spintronics view, a material possessing the following physical properties is promising: electrical conductivity and low stiffness. Electrical conductivity is especially required to replace the spin-wave spin current with a conventional spin current, which enables expansion of the material platform for spin mechanics studies. Low stiffness is also required to detect a small mechanical momentum due to the Einstein-de Haas effect. The YIG cantilever exhibits a considerably low frequency (~20 kHz) owing to its very long shape due to the very high elastic constants[5], which still limits the fusion of spin mechanics and microelectromechanical systems.

Hence, it is still quite significant to implement a search for a conductive and soft material in which spin current can be injected to observe the Einstein-de Haas effect. Here, we focus on Cu, which is conductive, is soft, and possesses a significantly long spin diffusion length at room temperature[6–8], and establish fabrication processes for a Cu spin valve with a suspended-channel structure. The number of studies on spin transport in a suspended channel is currently quite low, with graphene being the only example to the best of our knowledge.[9,10] Graphene is an atomically thin material, and thus, it is sensitive to impurities originating from a substrate. Spin transport in suspended graphene has been investigated by an electrical method and clarified to show a spin diffusion length longer than that of a conventional graphene channel due to the elimination of the effect from the substrate.[9] Meanwhile, spin transport in other materials with a suspended structure has not yet been reported, which is the other significance of our study.

**Results**

Figures 1(a)-(c) show the fabrication process of a Cu-based lateral spin valve (LSV) with a suspended channel. Square-shaped holes (2.0 × 1.8 μm$^2$ in size and 50 nm in depth) were fabricated on a SiO$_2$(300 nm)/Si substrate by using electron-beam (EB) lithography and reactive ion etching using CF$_4$ (the value in brackets indicates the thickness). These holes were filled with Al deposited by thermal heating evaporation prior to removing the resist. LSVs consisting of Cu(100 nm)/Ni$_{80}$Fe$_{20}$(25 nm) were fabricated using a lift-off process as the Cu channel was placed on the Al. Before evaporation of the Cu, the top of the Ni$_{80}$Fe$_{20}$ (Py) layer was etched with Ar-milling to remove residual resist to improve the quality of the Py/Cu interfaces. This process was performed *ex situ* of the Cu deposition chamber. Finally, a suspended Cu channel was formed by removing the Al beneath the Cu by using 2.38% tetramethyl-ammonium-hydroxide (TMAH) aq.[11] We named the LSVs with suspended channels Sus-LSVs. LSVs without suspended channels, i.e., conventional Cu spin valves, were also fabricated on the same substrate, named Fix-LSVs. Spin transport of the Fix-LSVs was measured before treatment with TMAH aq. Figure 1(d) shows an AFM image of a Sus-LSV. A LSV structure was formed as designed above the trench, and the heights of the Cu channel and Py electrodes above the trench were the same as those on the substrate. From the AFM observations, we corroborate that the Sus-LSV is successfully fabricated.

Figures 2(a) and (b) show the non-local 4-terminal magnetoresistance (MR) at room temperature of Fix- and Sus-LSVs when the centre-to-centre distance of the two Py electrodes ($L$) is set to 320 nm and 300 nm, respectively. Rectangular-shaped MRs were observed, which means that spin transport through the suspended Cu channel was achieved. $\Delta R_S = \Delta V/I$ was estimated to be 3.2 and 4.0 mΩ from Fix- and Sus-LSVs, respectively, where $\Delta V$ is the difference in the voltage under anti-parallel and parallel states of the Py electrodes and $I$ is the source current.

To obtain the spin diffusion length of the Cu channel, the $L$ dependence of MRs was measured (see Fig. 3). As mentioned in the sample preparation, the Py surface was exposed to air before evaporation of the Cu so that non-negligible interface resistance due to the natural oxide layer exists. Thus, the description of the $L$ dependence changes according to the size of the interface resistance of the Py/Cu interface and spin resistance of the Cu channel.[12] We measured the interface resistances $R_I$ of Fix- and Sus-LSVs by a 3-terminal measurement, and they were estimated to be 16.2 ± 9.4 Ω and 20.9 ± 16.4 Ω, respectively (note that the uncertainties were estimated as the standard errors from the measurement of devices with different values of $L$). We also estimated the resistivity of our Cu channel in the Fix- and Sus-LSVs to be $2.92 \times 10^{-8}$ Ω·m via a conventional 4-terminal measurement. Hence, the spin resistances of Cu channels $R_{SN}$ were estimated to be 0.38 ± 0.04 Ω in the Fix-LSV and 0.35 ± 0.03 Ω in the Sus-LSV. Here, $R_{SN}$ is described as $R_{SN} = \rho\lambda/A$, where $\rho$ is the resistivity, $\lambda$ is the spin diffusion length, and $A$ is the cross-sectional area of the channel.[12] We assumed a spin diffusion length of Cu $\lambda_{Cu}$ = 350 nm at room temperature[6], and $A$ was obtained from SEM images for each device (not shown). Since $R_I$ was much larger than $R_{SN}$, $\Delta R$ could be described as follows[12]:

$$\Delta R_S = P^2 R_{SN} e^{-\frac{L}{\lambda_{Cu}}}. \tag{1}$$

Here, $P$ is the spin polarization at the Cu/Py interface. The natural logarithm can be taken on both sides of Eq. (1), yielding:

$$\ln(\Delta R_S) = -\frac{L}{\lambda_{Cu}} + \ln(P^2 R_{SN}). \tag{2}$$

Then, we estimated $\lambda_{Cu}$ to be 390 ± 40 nm in the Fix-LSV and 340 ± 40 nm in the Sus-LSV from the slope of the linear relationship between $\ln(\Delta R_S)$ and $L$, as shown in Fig. 3(b). The solid lines in Fig. 3(a) show the fitting lines relative to Eq. (1), with $\lambda_{Cu}$ were set to the values estimated above.

**Discussion**

The spin diffusion length of the Sus-LSV was slightly shorter than that of the Fix-LSV, and the suspended structure did not seem to influence the spin transport in Cu. Even if we consider the small difference in the spin diffusion lengths, it is attributable to the formation of an oxide layer at the bottom of the Cu channel in the Sus-LSV. The naturally oxidized surface of Cu can be an origin of spin scattering, as experiments on the thickness dependence of the spin diffusion length of a Cu channel have shown.[8] For the Sus-LSV, the bottom side was exposed to air, resulting in oxidation; hence, the spin scattering probability in a Sus-LSV can be enhanced because spin-scattering enhancement takes place on the bottom surface of a Sus-LSV. Meanwhile, when $L$ is shorter than 400 nm, the $\Delta R$ in the Sus-LSVs is larger than that in the Fix-LSVs. This increase is attributed to the enhancement of $R_{SN}$ of the Sus-LSV when we assume $P$ is the same in both the Fix-LSV and the Sus-LSV. The ratio of $R_{SN}$ in the Sus-LSV to that in the Fix-LSV is estimated to be 1.94 ± 0.78 from the intercepts of the linear fittings of Fig. 3(b), which can be explained by the existence of oxidized Cu. Since naturally oxidized Cu in air is a semiconductor (cuprous oxide, $Cu_2O$) and highly resistive[13], the effective cross section, $A$, of the Sus-LSVs can be reduced, resulting in an enhancement in $R_{SN}$. In addition, TMAH aq., used to remove the Al layer in the sample fabrication, may also dissolve the oxidized Cu layer, since previous studies showed that alkali aqueous solutions including ammonia help etch oxide Cu.[14–16] This also suggests a decrease in the effective $A$ in the Sus-LSVs via the re-oxidization of Cu in air (note that the Fix-LSVs did not undergo the TMAH aq. treatment). The thickness of the Cu channel was measured with AFM, and it was found to decrease by 2-5 nm after treatment with TMAH aq., which is a reasonable thickness expected from naturally oxidized Cu in air.[13] When we assume that $A$ is reduced after a 5 nm-thick oxidized Cu layer in the surface is dissolved, the ratio of $A$ values is estimated to be 1.2, which is in the range of the ratio of $R_{SN}$ values estimated by the experimental results. Thus, $\Delta R_S$ in the Sus-LSVs can be larger than that in the Fix-LSVs only due to the oxidization of the Cu surface.

In conclusion, we established a method to fabricate LSVs with suspended Cu channels. This technique is applicable to other spin-current channels. Moreover, we successfully observed spin transport through a channel via non-local 4-terminal measurements. The spin diffusion length was estimated to be 340 nm from the channel-length dependence of the signal amplitude. The small difference between spin diffusion lengths in the fixed and suspended Cu channels is attributed to the oxidization of the Cu on the bottom surface and not to the suspended structure. This result indicates that suspended Cu can be useful for creating cantilevers with a sufficient size to study spin mechanics.

**Methods**

**Sample fabrication**

Square-shaped holes were formed with electron-beam (EB) lithography and reactive ion etching (RIE). The resist for EB lithography was ZEP-520A (Zeon Corporation). RIE was performed for 3 minutes with CF4, where the gas flow was 50 sccm, the pressure was 5.0 Pa, and the RF power was 50 W. These holes were filled with Al deposited by resistance heating evaporation. Py electrodes were prepared with EB lithography and EB deposition. The Cu channel was also prepared with EB lithography and resistance heating evaporation. Before Cu evaporation, the tops of the Py electrodes were etched with Ar-milling *ex situ* to remove a residual resist. After fabrication of the LSV structure, the Al was removed from the bottom of the channel by soaking the sample in 2.38% TMAH aq. for 90 seconds (NMD-3, Kanto Chemical Industry Co., Ltd.).

**Measurement**

For the non-local 4-terminal measurement, current $I$ was injected from one of Py electrode with a current source (SourceMeter, Keithley 2400), and the voltage on the other Py electrode was detected with a nanovoltmeter (Keithley 2182A), as shown in Fig. 1(d). $I$ was set to 300 μA. A magnetic field $|H| \leq 300$ Oe was applied along the long axis direction of the Py electrodes. To estimate the interface resistance of the Cu/Py interface, we measured the current-voltage characteristics in a conventional 3-terminal measurement setup with SourceMeter and a nanovoltmeter connected to opposite sides of the same electrodes. The voltage ranged from ±400 mV, with a step of 40 mV. Every measurement was carried out at room temperature.

**ACKNOWLEDGEMENTS**

This research was supported in part by a Grant-in-Aid for Scientific Research from the Ministry of Education, Culture, Sports, Science and Technology (MEXT) of Japan, Scientific Research (S) "Semiconductor Spincurrentronics" (No. 16H0633).

**Figure captions**

Figure 1: Fabrication process of the suspended Cu channel: (a) make a hole and refill it with Al, (b) fabricate a LSV on top of the previously deposited Al, and (c) remove the Al via TMAH aq. (d) An AFM image of the LSV with the suspended channel and the measurement setup. The scan range was 5 × 5 μm². A magnetic field $H$ was applied along the long side direction of the Py electrode. The current was injected from one of the Py electrodes, and the voltage was measured with the other electrode.

Figure 2: Magnetoresistance in the non-local 4-terminal measurement with (a) a Fix-LSV (the centre-to-centre distance of Py electrodes $L$ = 320 nm) and (b) a Sus-LSV ($L$ = 300 nm).

Figure 3: (a) $L$ dependence of $\Delta R_S$ obtained from the Fix-LSV and the Sus-LSV. Here, $L$ is the centre-to-centre distance of the Py electrodes, and $\Delta R_S$ is described as $\Delta V/I$, where $\Delta V$ is the difference in voltage between the anti-parallel and parallel states of the Py electrodes and $I$ is the source current. The solid lines are obtained from Eq. (1) in the main text, with $\lambda_{Cu}$ = 390 nm in the Fix-LSV and $\lambda_{Cu}$ = 340 nm in the Sus-LSV. (b) Natural logarithm of (a) on the vertical axis. The solid lines are obtained from Eq. (2) in the main text. The error bars represent the standard errors obtained from the parallel-state voltage in Fig. 2.

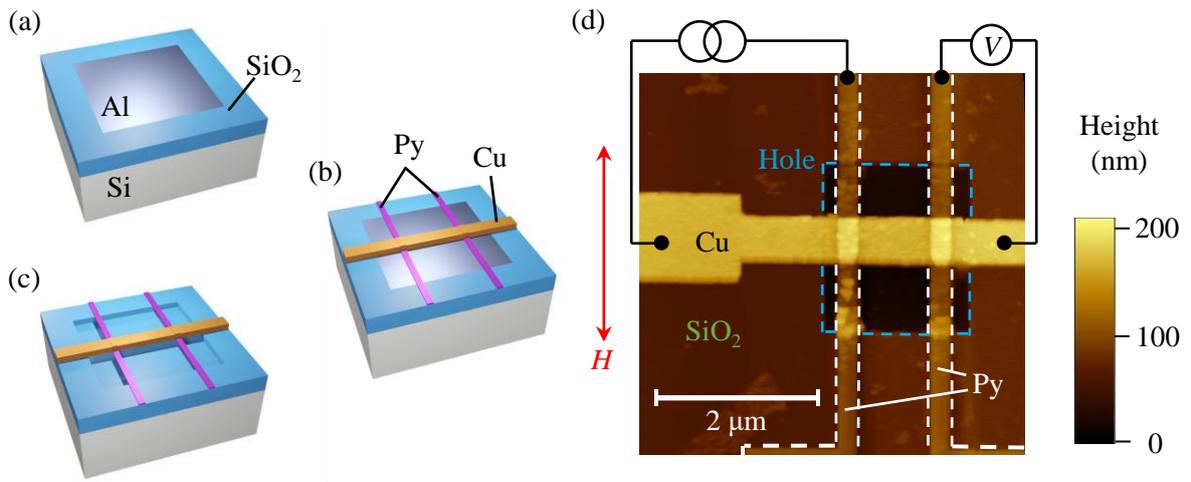

Fig. 1  K. Matsuki *et al.*

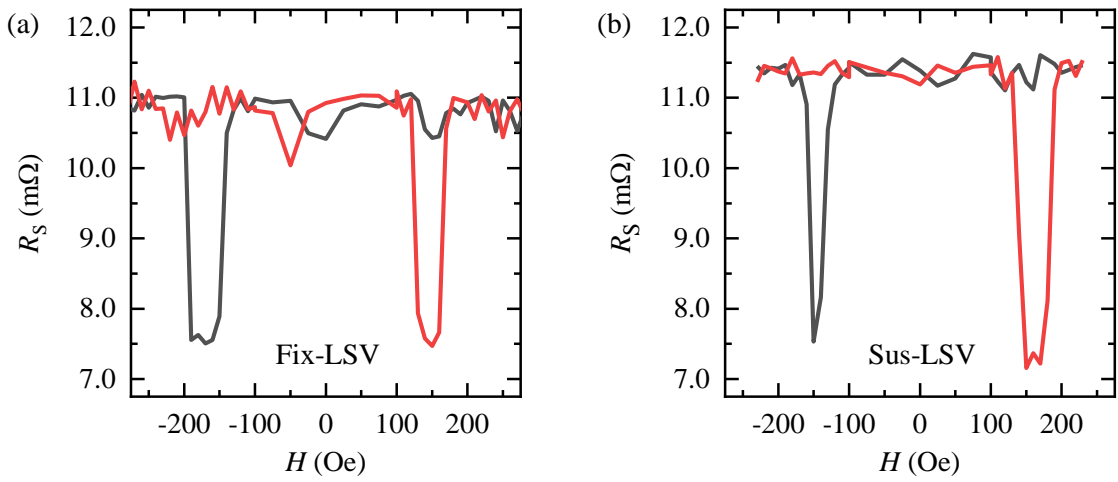

Fig. 2  K. Matsuki *et al.*

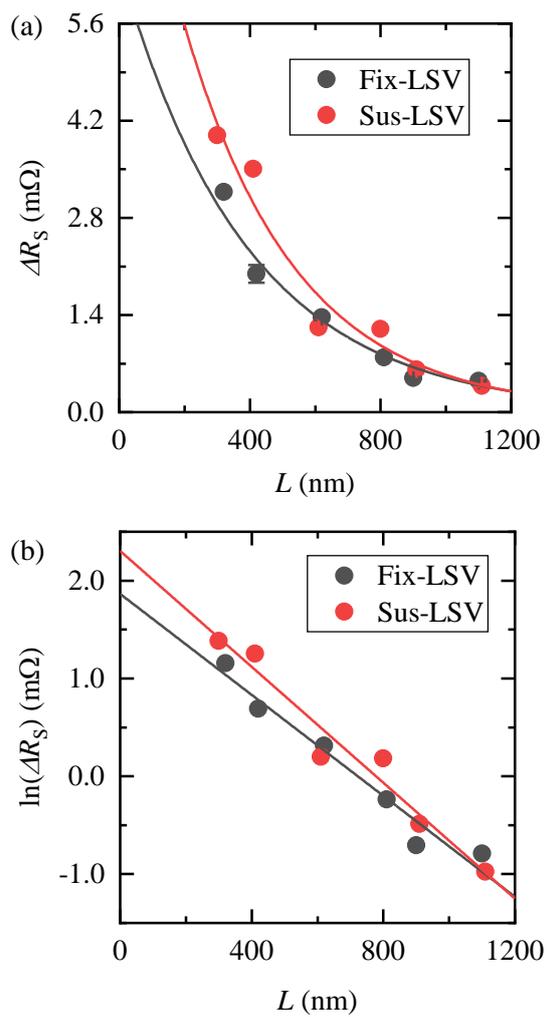

Fig. 3  K. Matsuki *et al*.